\documentclass[%
 reprint,
superscriptaddress,
nofootinbib,
 amsmath,amssymb,
 aps,
]{revtex4-1}

\usepackage{xcolor}
\usepackage{graphicx}
\usepackage{dcolumn}
\usepackage{bm}

\newcommand{\xmax}{$\langle X_{\rm max} \rangle$}
\newcommand{\sx}{$\sigma_{X_{\rm max}}$}
\newcommand{\apjl}{ApJL}
\newcommand{\jcap}{JCAP}

\begin{document}


\title{What do the highest-energy cosmic-ray data suggest\\  about
  possible new physics around 50 TeV?}

\author{Vasiliki Pavlidou }
\affiliation{
Department of Physics and 
Institute for Theoretical and Computational Physics, 
University of Crete}
\affiliation{
Foundation for Research and Technology - Hellas, IESL, 71110,
Heraklion, Greece }
\author{Theodore Tomaras}
\affiliation{
Department of Physics and 
Institute for Theoretical and Computational Physics, 
University of Crete}
\date{\today}
\begin{abstract}
The latest observations of extensive air showers (EAS) induced by
ultra-high-energy cosmic rays (UHECR) appear to indicate, {\it prima
  facie}, a transition to heavy 
primaries at the highest energies. 
However, this interpretation, which is based on
extrapolations of the Standard Model (SM) to ultra-LHC energies, is
strained from both astrophysical and particle phenomenology
perspectives.  
We consider the alternative that after some
energy threshold, the first collision of the primary in the atmosphere 
results
 in a state, the decay of
which leads to a considerably increased shower particle multiplicity, 
so that light-primary EAS appear
heavy-like. We show that a minimal
implementation of such a model yields
predictions for the average EAS depth and  shower-to-shower
fluctuations 
that are consistent with
each other, and an excellent fit to Auger data. If such an effect
indeed takes place, 
we predict that: (a) the center-of-momentum (CM) energy
threshold  for the effect is of order 50 TeV;
(b) the probability with which the effect occurs is
high, and it will be detected easily by next-generation
accelerators; (c) the increase in multiplicity
compared to the SM prediction 
grows with CM energy roughly as $\sim
E_{\rm CM}$;
(d) the cosmic-ray composition at the highest energies is
light. Remarkably, if the latter 
is confirmed electromagnetically
this would
necessitate the existence of new physics by these energies.
\end{abstract}

\maketitle

{\bf Introduction.}~Ultra-high-energy cosmic rays (UHECR) are the
highest-energy particles in the 
Universe. They are extremely rare (one
particle per ${\rm km^2}$ per year at energies above $10^{18} {\rm \,
    eV}$). Even so, thanks to the operation of cosmic-ray observatories
spanning  thousands of km$^2$, there has
been, in the past fifteen years, an explosion of 
unprecedented-quality data \cite{hires-spectrum,
  auger-spectrum,auger-xmax, ta-spectrum}. Results from
HiRes \cite{hires-observatory}, the Pierre Auger Observatory
\cite{auger-observatory}, and Telescope Array \cite{ta-observatory},
now allow the use of UHECR as probes of high-energy physics. 
The largest cumulative exposure at the highest energies
($>6.7\times10^{4} {\, \rm km^2 \, sr \, yr}$, \cite{auger-icrc2017-highlights}) has been
achieved by the Auger Observatory, 
and it
is the interpretation of the latest Auger data above $10^{17.5} \rm \,
eV$ \cite{auger-icrc2017} that we 
focus on. 

This plethora of high-quality data has exposed new
puzzles in cosmic-ray physics. The most
pressing one involves the composition of UHECR and its evolution with
energy.
 All composition-sensitive observables appear to indicate,
{\it prima facie}, that, at the highest energies, heavier nuclei start
to dominate over protons \cite{auger-xmax, auger-composition,
  auger-compfit}; however the results from these observables are not
fully consistent with each other \cite{auger-icrc2017}. 

\begin{figure}
\resizebox{1.0\hsize}{!}{\includegraphics[scale=0.7,clip]{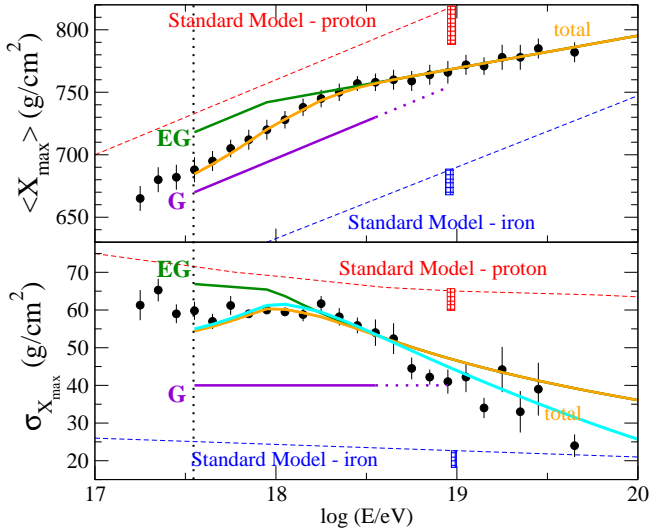} }
\caption{Upper panel: \xmax \ as a function of energy. Filled circles:
  Auger 2017 ICRC data (error bars are systematic).  Red/blue dashed lines:
  SM (Sibyll) predictions for protons/iron, from
  \cite{auger-icrc2017-highlights}. The hatched boxes indicate the
  systematic uncertainty of SM predictions (result of using
  EPOS/QGSJet instead of Sibyll).  Thick lines: our model (purple: Galactic; green:
  extragalactic; orange: total). Lower panel: $\sigma_{X_{\rm max}}$
  as a function of energy. Filled circles: Auger ICRC  2017 data
  (error bars are statistical). Orange line: $\delta =0$.Cyan line:
  $\delta = 2.9$. Other lines as above. For clarity, the extragalactic
  model is only shown for $\delta =0$.}
\label{fig:xmax}
\end{figure}

The distribution, in a given primary energy range, of the atmospheric slant
depth $X_{\rm max}$ (expressed as column density) where the 
energy deposition rate of EAS particles in the atmosphere reaches its
maximum value is both  composition-sensitive
\cite{elongation-rate-theorem,ku12}, and directly observable by  
fluorescence detectors. For this reason, its first two moments
(average shower depth, \xmax, and standard deviation, \sx) are the
most widely used composition-sensitive observables. 
Auger data on both \xmax \ and
\sx \ show a qualitative trend towards 
heavy-like EAS above $\sim 2\times 10^{18}$ eV (see
Fig.~\ref{fig:xmax}), however the two datasets are 
not straightforward to reconcile in detail, with the Auger Collaboration
reporting strained fits to the observed $X_{\rm max}$ distribution
in more energy bins than what expected from random fluctuations alone:
there is {\em no}  primary composition that can fully reproduce the
observed distributions \cite{auger-icrc2017}. Additional
composition-sensitive quantities  
obtained from the surface water-Cherenkov
detectors, when interpreted using SM EAS simulations,
yield a mass composition heavier than the one derived
from $X_{\rm max}$, with the discrepancy traced
to an observed excess of muons compared to
SM expectations \cite{auger-icrc2017}.  
This is not surprising, as the interpretation of
composition-sensitive observables relies on simulations of EAS 
development, which in
turn draw on extrapolations of SM results to ultra-LHC
energies.

 The alternative, therefore, to the UHECR composition getting heavier,
is that there is some new physical
effect, yet-unseen in accelerators, that takes place in the first
collision of UHECR primaries in 
the atmosphere above some energy threshold $E_{\rm th}$ and affects the shower development.
That this scenario is an open possibility is widely recognized by the Auger
Collaboration (e.g., \cite{auger-icrc2017, auger-compfit,
  auger_muon_excess}) and other 
authors (e.g., \cite{FA,fireball,LI_violation}). Here, we quantify
phenomenological constraints encoded in Auger data for
any new phenomenon that could be affecting EAS development. 

Specifically, assuming that, at energies $>2\times10^{18}\,{\rm eV}$:\\
(a) a
single population of extragalactic cosmic rays dominates; \\
(b) the composition of extragalactic cosmic rays remains
light;\\
 (c) the  - abnormal for
protons and light nuclei -  growth of \xmax \ with energy reflects the
phenomenology of this new physical effect, \\ 
we show that Auger data on \xmax \ and  \sx \ can be readily
reproduced. 

{\bf What kind of new physics?} The primary requirement for a
candidate new physical effect is
to make light-primary EAS appear ``heavy-like'', which in practice
translates to (a) having a smaller \xmax \ and (b) having smaller \sx
\ than the SM prediction
for protons. 

The phenomenology we consider is that the first  collision of the primary
in the atmosphere  results, with high probability,  in a state the
decay of which leads to a considerably increased particle 
multiplicity early in the shower. A large number of particles injected
early in the shower development will lead to showers that reach their
maximum at smaller values of $X$, as well as smaller $\sigma_{X_{\rm
    max}}$ (as shower-to-shower 
fluctuations will average out).  
 
Several candidate particles and new physics
mechanisms that might lead to such a behavior are reviewed in
\cite{MMT,MS}. They are based 
either on the possible existence of yet undiscovered particles (mini
black holes, “strangelets”) or on special phases of QCD, such as the
disoriented chiral condensate (DCC). The mini black hole paradigm has
been analyzed  in detail in \cite{cct}, while a recent proposal based
on chiral symmetry 
restoration in QCD can be found in \cite{FA}.  

The
quantitative impact of such a scenario on composition-sensitive
observables is model-dependent; a rough 
phenomenological estimate is however straightforward to make.

{\bf Growth of \xmax \ with energy.} For a single shower, $X_{\rm
  max} = X_1+X_{\rm D}$, with $X_1$ being the depth of the first interaction
and $X_{\rm D}$ being the additional column density required for the shower to reach
its maximum development. For energies below $E_{\rm th}$, SM
predictions hold. $\langle X_1 \rangle = m/\sigma_{\rm p-air}$ where
$m$ is the average atomic mass of air ($\simeq 14.5$ proton masses,
e.g. \cite{ueu11})
and $\sigma_{\rm p-air}$ is the proton-air cross section\footnote{We
use the ${\rm Sibyll \, 2.1}$ extrapolation $\sigma_{\rm p-air} \simeq 520 {\rm
  \, mb} + 60 {\rm \, mb} \log (E/10^{17.5} {\rm \, eV})$ \citep{ueu11};
our results are not sensitive to this choice.}. We parameterize
$\sigma_{\rm p-air} \simeq \sigma_0+\beta \log\epsilon$ for $\epsilon
\leq 1$, where
$\epsilon = E/E_{\rm th}$. Any new phenomenon will likely affect
$\sigma_{\rm p-air}$, so that $\sigma_{\rm p-air} \simeq \sigma_0+\beta' \log\epsilon$ for $\epsilon
\geq 1$, assuming that $\sigma_{\rm p-air}$ is continuous as the slope
changes\footnote{More generally, $\sigma_{\rm p-air}$ might also
  exhibit a discontinuity at $\epsilon =1$. For simplicity, we do not
  make use of this extra freedom. }
 from its SM value $\beta$ to $\beta'$. Thus, for $\epsilon
\geq 1$, $\langle X_1
\rangle \simeq (m/\sigma_0) - (m\beta'/\sigma_0^2)\log
\epsilon$. 

 The change in $X_{\rm D}$ is entirely due to an  increase in partcile 
 multiplicity at the first collision, since the products will have, on
average, energies below $E_{\rm th}$. We parameterize the change in
multiplicity by 
$n(\epsilon) \equiv N(\epsilon)/N_{SM}(\epsilon) > 1$ (for $\epsilon
\ge 1$), where
$N(\epsilon)$ and $N_{SM} (\epsilon)$ are the actual and  SM-predicted (by
shower simulations) number of first collision products. 
We can then empirically model the
shower as $n(\epsilon)$ ``component-showers'' (CS) of
energy, on average, $\epsilon/n(\epsilon)$, developing independently. 
Since for $\epsilon \leq 1$ the
SM prediction \cite{auger-icrc2017} is $\langle X_{\rm D}
\rangle \simeq \langle X_{\rm D} \rangle (1)+(65 {\rm g/cm^2} )\log
\epsilon$, for $\epsilon \geq 1$ we obtain    $\langle X_{\rm D}
\rangle \simeq \langle X_{\rm D} \rangle (1)+(65 {\rm g/cm^2} )\log
\epsilon/n(\epsilon)$ (where we have assumed $n(1)=1$). 

 The Auger Collaboration \cite{auger-icrc2017} fits, for $E\gtrsim 2\times 10^{18} \, {\rm eV}$,
$\langle X_{\rm max}\rangle /{\rm g \, cm^{-2}} \sim (26 \pm2) \log
\epsilon$. In the simplest case where
the new state is produced almost in every EAS for $\epsilon \geq 1$,
  assuming that the
composition at these energies 
  remains constant, and the difference
with the SM prediction is purely due to new physics, we can obtain
$n(\epsilon)$ by demanding
that, \begin{equation}
65 \log [\epsilon/n] - \frac{m \beta'}{\sigma_0^2}\log\epsilon =
26 \log \epsilon\,.
\end{equation}
This yields\begin{equation}\label{our_n}
n(\epsilon) \simeq \epsilon^{0.52 -0.08\delta}\,,
\end{equation}
where $\delta = \beta'/\beta-1$.

{\bf Change of \sx \ with energy}. The $X_{\rm max}$ spread between showers
is the joint effect of fluctuations in $X_1$ and in shower
development, $\sigma^2_{X_{\rm max}}=\sigma^2_{X_1} + \sigma_{X_{\rm D}}^2$, with
$\sigma_{X_1} = \langle X_1\rangle$ (Poisson statistics). 
 To estimate $\sigma_{X_{\rm D}}$, we
take the average $(1/n)\sum_i X_{{\rm D},i}$ of individual CS
maxima to be a reasonable estimator of the overall $X_{\rm D}$. Then
$X_{\rm D}$
is the ``sample mean'' of $n$ ``draws'' from the underlying
distribution of $X_{\rm D,i}$, and the distribution of these
``sample means'' has a spread that is given by the ``error in
the mean'' formula, $\sigma_{X_{\rm D}} =  \sigma_{X_{{\rm D},i}} / \sqrt{n}$. Here 
$\sigma_{X_{{\rm D},i}}$ is the spread of $X_{{\rm D,}i}$, and it can
be assumed to follow the SM predictions, since the individual
energies of
the decay products initiating the CS are $<E_{\rm th}$. The SM
predicts that $\sigma_{X_{{\rm D},i}}$ is approximately constant
(the mild decline with energy predicted by SM shower simulations for \sx in the case
of protons can be reproduced by the logarithmic rise of
$\sigma_{\rm p-air}$ with energy). Therefore
\begin{equation}\label{our-sigma}
\sigma^2_{\rm X_{\rm max}} (\epsilon)= 
\sigma^2_{\rm X_1}(1) - 
10.7 \frac{\rm g}{\rm cm^2}\sigma_{\rm X_1} (1) (1+\delta)\log\epsilon
+
\frac{\sigma^2_{\rm X_{D}}(1)}{n(\epsilon)}.
\end{equation}

{\bf A proof-of-principle minimal model.} As a proof of principle for
this concept, we show how a simple two-component  
astrophysical scenario (heavy Galactic cosmic rays cutting off; light
extragalactic cosmic rays dominating at high energies) with EAS
obeying Eqs.~(\ref{our_n}) and (\ref{our-sigma}) above $E_{\rm th}$
reproduces well Auger data on \xmax, 
\sx, and yields reasonable flux spectra for the two populations. 

For a mixture of Galactic and extragalactic cosmic rays with a
fraction of Galactic over total particles $f(\epsilon)$,
the probability density function of $X_{\rm max}$ will be $p(X_{\rm max})=f p_G(X_{\rm max})+(1-f)p_{EG}(X_{\rm max})$, so that
 \xmax  \ will be given by 
\begin{equation} \label{xmaxmix}
\langle X_{\rm max} \rangle = f \langle X_{\rm max} \rangle_G +
(1-f)\langle X_{\rm max} \rangle_{EG}\,,
\end{equation} 
and $\sigma_{X_{\rm max}}^2$ by 
\begin{eqnarray}
\sigma_{X_{\rm max}}^2  &=& f \sigma_{X_{\rm max}, G} ^2  +
                            (1-f)\sigma_{X_{\rm max},EG}^2
                  \nonumber \\
&&+f(1-f) \left( \langle X_{\rm max} \rangle_G - \langle
   X_{\rm max} 
\rangle_{EG} \right)^2\, \label{sxmix}
\end{eqnarray}
with subscripts $G$ and $EG$ referring to the Galactic and extragalactic
populations respectively. 

There is little freedom in this model. Assuming that extragalactic
cosmic rays have completely dominated for $E > 2 \times 10^{18}$ eV,
the evolution of \xmax$_{\rm EG}$ can be directly read off of the Auger
data in this energy range, 
$\langle X_{\rm max} \rangle_{\rm EG}  / {\rm g \, cm^{-2}}=728 + 26 
  \log(\epsilon/\epsilon_{17.5})\,,$ where $\epsilon_{17.5}= 10^{17.5}
  {\rm eV} /E_{\rm th}$. The
  continuity assumption for $n(\epsilon)$, and, consequently for \xmax$_{
    EG}(\epsilon)$ then fully determines 
the behavior of \xmax$_{\rm EG}$ at Auger energies, if the value of
$E_{\rm th}$
is known. 

\begin{figure}
\resizebox{1.0 \hsize}{!}{\includegraphics[scale=0.6, clip]{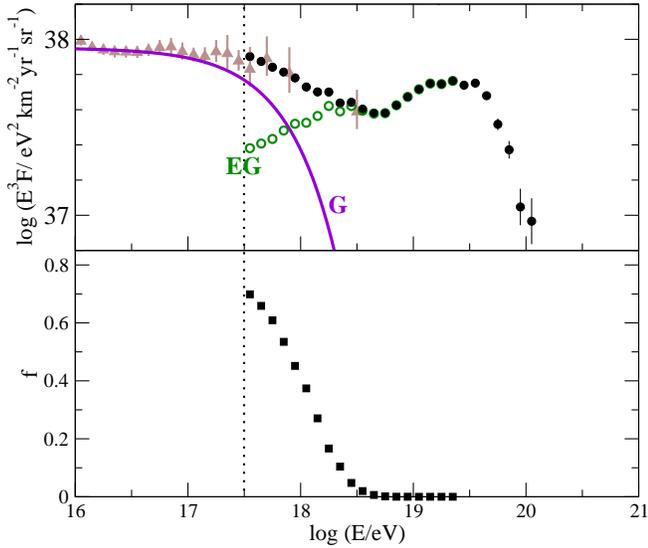} }
\caption{Upper panel: cosmic ray spectrum between $10^{16}$ and
  $10^{20}$ eV. Filled circles: Auger 2017 ICRC spectrum \cite {auger-icrc2017} (error bars are
  statistcical). Brown triangles: KASCADE-Grande 2015 all-particle
  spectrum \cite{kg-icrc2015}, QGSJET II - 04  reconstruction (error bars are
  systematic). Purple line: Galactic population model spectrum (this
  work). Open green circles: Auger total flux minus Galactic model. The
  vertical black dotted line indicates the lowest energy for which
  there are spectrum measurements from Auger. Lower panel: fraction of cosmic rays of
  Galactic origin as a function of energy, derived from the Galactic
  flux model over the total observed flux. }
\label{fig:spectrum}
\end{figure}

A similarly strong statement can be made for $f$. The shape of the
extragalactic  
population flux spectrum is affected by intergalactic losses (which in turn 
depend on the composition of extragalactic cosmic
rays, the distribution and cosmic evolution of extragalactic 
cosmic-ray sources, and the cosmic density of diffuse photon
backgrounds) and the pileup of particles 
down-cascading from higher energies 
\cite{allard2006,allard2008,kotera2010,aloisio2013,aloisio2015}.  These 
are non-trivial to calculate theoretically, because of the
uncertainties involved in the inputs, but also because any systematic
uncertainties in the energy reconstruction of cosmic-ray events
shift the energy location where specific absorption features appear. 
In contrast, the Galactic cosmic-ray flux is reasonably
expected to be a declining power law (from Fermi acceleration) with an
exponential cutoff (induced by Galactic accelerators reaching the
maximum energy they can achieve), 
$F_{\rm G}(\epsilon) = F_{\rm
  G,0}(\epsilon/\epsilon_{17.5})^{-\gamma_{\rm G}}
\exp\left[-\epsilon/\epsilon_{\rm G}\right]$. 
The values of $F_{\rm G,0}$ and $\gamma_{\rm G}$ are well-constrained by
KASCADE-Grande data at lower energies\footnote{
We adopt purely empirically, the 2015 ICRC QGSJetII-04 --based energy
  reconstruction of KASCADE-Grande events \cite {kg-icrc2015}, which
 results in a  
 near-perfect continuity with Auger measurements at overlapping
  energies, see Fig.~\ref{fig:spectrum}. 
}, with $F_{\rm G,0}\simeq 2 \times
10^{-15}$ $ {\rm \, km^{-2} yr^{-1} sr^{-1}eV^{-1}}$ and
$\gamma_{\rm G} \simeq 3$ (see Fig.~\ref{fig:spectrum}). The value of $\epsilon_{\rm G} = E_{\rm
  G}/E_{\rm th}$ can then be constrained by
the requirement that the flux residuals $F_{\rm total, Auger} (\epsilon)-
F_{\rm G}(\epsilon)$ in the lower-energy part of the Auger range, before any
intergalactic propagation losses set in, are consistent with a power
law (again assuming Fermi acceleration for extragalactic sources). For
values outside the range $6.5 \times 10^{17} {\rm eV} <E_{\rm G}< 8.5
\times 10^{17} {\rm eV}$ the 
low-energy Auger residuals (see Fig.~\ref{fig:spectrum}, upper panel,
green open circles) start to
exhibit curvature in a log-log plot. We adopt $E_{\rm G} = 7.5 \times
10^{17} {\rm eV}$, in the 
middle of this range (purple line, 
Fig.~\ref{fig:spectrum}, upper panel). 
 This then fixes $f(\epsilon)$ to $ F_{\rm G}(\epsilon)/F_{\rm
 total, Auger}(\epsilon)$ (Fig.~\ref{fig:spectrum}, lower panel). 
 
The Galactic component is heavy.  The exact composition is
subject to various systematic uncertainties  
\cite{kg_comp_2015,kg-icrc2015}, so for simplicity, we take the SM
predictions for carbon nuclei 
(\xmax$_{\rm G,0} \simeq 670 {\rm g/cm^2}$ and \sx$_{\rm G,0} \simeq 38 {\rm
  g/cm^2}$ at $10^{17.5} {\rm \, 
  eV}$, from a naive extrapolation of data presented in 
\cite{multielement,auger-icrc2017}) to be representative, on average, of the
behavior of EAS initiated by Galactic cosmic rays\footnote{The
  composition of Galactic cosmic rays evolves strongly between the
  knee ($\simeq 10^{15.5} {\rm \, eV}$) and their final cutoff at $E_{\rm G}$. Our simple assumption cannot capture this behavior and thus we do not expect to fit the data below $10^{17.5} {\rm \, eV}$.}. We have however
verified that more complex mixes also give good fits with other model
inputs within their respective allowed ranges. Since \sx \ evolves
very little for heavier nuclei in the energy range relevant for the
Galactic population, we take it to be constant for simplicity. Because
$f(\epsilon)$ is highly suppressed by the energy new physics sets in,
these choices affect neither our fit to Auger data at the high
end of their energy range, nor our conclusions on possible new physics
phenomenology. 

For both a pure proton population and any reasonable light mix,
\sx$_{\rm EG,0}$ will
be $68 \, \pm $ a few ${\rm g/cm^2}$ at $10^{17.5}{\rm \, eV}$
\cite{auger-icrc2017}.
 We take  \sx$_{\rm EG,0} = 68 {\rm \, g/cm^2}$.

A nominally free parameter in our model is the threshold
energy, $E_{\rm th}$, where new physics sets in. However its value is
very well bounded.  By the requirement that \xmax$_{\rm EG}$ does
not, at any energy, exceed (within
systematic uncertainties) the SM predictions for protons, $E_{\rm th}
\gtrsim 10^{17.5} \rm \, eV$ (see Fig.~\ref{fig:xmax}, upper panel). This
corresponds to  
$E_{\rm CM,th}\gtrsim 25 \, {\rm TeV}$, in
agreement with the non-detection by the LHC of any effects deviating
from SM predictions. By the assumption that new physics
has already set in by the break observed by Auger in \xmax, 
$E_{\rm th} \lesssim 10^{18.3} {\rm \, eV}$. Good fits to the Auger
dataset can be obtained throughout this narrow range, given the
uncertainties in the Auger data and the allowed range in other
model inputs. In what follows, we will use  $E_{\rm th} \simeq 10^{18}
{\, \rm
  eV}$ ($E_{\rm CM,th}\simeq 45 \, {\rm TeV}$). 
For heavier primary nuclei, the per-nucleon threshold for mass number A is reached
at a higher primary  energy, $AE_{\rm th}$. For this reason, the new
physics never becomes 
relevant for Galactic cosmic rays, as extragalactic cosmic rays have
completely dominated before $AE_{\rm th}$ is reached, for any reasonable
$A$ (hence the ``agnostic'' dotted lines for the
Galactic population at high energies in Fig.~\ref{fig:xmax}). 

This leaves a single free parameter in our model, $\delta$, which
affects $X_1$. \xmax \ shows no sensitivity to $\delta$, because it is
dominated by  $\langle X_{\rm D} \rangle$. In contrast \sx \ is more
sensitive to $\delta$; however, at the high energies where its effect
becomes important, Auger \sx \ data have large statistical uncertainties. In
Fig.~\ref{fig:xmax}, we show two cases: $\delta = 0$
($\sigma_{\rm p-air}$ is not affected by new physics, orange line),
and $\delta = 2.9$ (cyan line). Note that even the latter case is consistent with SM
predictions within uncertainties \cite{ueu11}.

{\bf Results and Discussion.} The resulting \xmax$(E)$ and \sx$(E)$ curves are
shown in 
Fig.~\ref{fig:xmax}. 
In the same energy range, the two datasets
resemble broken logarithmic 
growth with two different 
slopes; the Auger Collaboration fits
them as such \cite{auger-icrc2017}. Each such relation involves four free
parameters, so fitting the two datasets in this
way would require eight free parameters. We have incorporated in our
model the slope and normalization of the second branch of \xmax, so
a purely empirical model would need another six free
parameters to fit both datasets well. Without using {\em any} of this
freedom, we have produced model curves
for two very different values of $\delta$ 
that perform better than Astrophysical scenarios (extragalactic
accelerator composition getting heavier)
\cite{auger-compfit,khota-jets,ke-khota,aloisio2015, gap-complete};
and all other inputs in our 
model are driven by astrophysics and/or the requirement of 
consistency with the  SM predictions at low energies.  

In addition, astrophysical scenarios with a transition to
heavier composition at the highest energies generally do not 
attempt to reproduce the entire Auger energy range (e.g.,
\cite{auger-compfit,khota-jets,ke-khota,aloisio2015}), but focus
instead above $\sim 5\times 10^{18}$ eV, leaving room for a possible third 
component between Galactic cosmic rays and the highest-energy cosmic
rays, an issue explicitly addressed by \cite{auger-compfit} (see
however \cite{gap-complete,ufa15} for  models that treat the entire Auger
energy range). 

Astrophysical explanations
of the shallow growth of \xmax \ 
at the highest energies  have to invoke two ``cosmic coincidences'':
(a) the Galactic/extragalactic accelerator coincidence at $10^{18.5}
{\rm \, eV}$: the energy 
where the Galactic accelerators cut off is close to the energy where the
composition of extragalactic accelerators starts getting heavier; 
(b) the extragalactic accelerator / cosmic
photon background coincidence at  $10^{19.5} {\rm \, eV}$: the maximum energy 
achievable by extragalactic accelerators is  close to the energy threshold for
photopion/photodissociation energy losses (the Greisen - Zatsepin -
Kuzmin, GZK, cutoff \cite{gzk-g,gzk-zk}).  Neither issue appears in our
scenario, where extragalactic accelerators remain efficient and their
output light throughout the
Auger energy range. 
In our scenario, the energy scale of $2 \times 10^{18}$ eV where the
slopes of \xmax \ and \sx \ are seen to change in the data {\em does not}
represent the energy where new physics sets in; rather, this break is
astrophysical, and signifies 
extragalactic cosmic rays dominating over the Galactic
population. The new effect has already appeared at a lower
energy. 

Our empirical model does not treat the muon excess; 
we note however that  both production of mini black holes and the
restoration of chiral symmetry paradigms might in principle alleviate
the muon deficit problem.
The simple implementation of the new effect we have presented here is
only meant as a proof of principle.
Ultimately, the impact of specific models on
EAS phenomenology, including their ability to alleviate the muon
excess, can be best
studied using EAS simulations as, e.g,
in \cite{cct, FA}.

The phenomenology we have considered here leads to {\bf four specific
predictions} with important implications for future astroparticle and
particle physics experiments.

\begin{enumerate}

\item The increase in multiplicity relative to the SM, $n(E)$, grows
  with lab-frame primary
energy as $\sim E^{0.52 -0.08\delta}$
(and with CM energy as $E_{\rm CM}^{1.04 -0.16\delta}$). 
Curiously, 
the multiplicity of the decay of mini black holes depends on the black
hole mass $M_{\rm BH}\propto E_{\rm CM}$ as $M_{\rm BH}^{(n+2)/(n+1)}$ (where $n$ is the
number of extra dimensions), in general
agreement with the empirical relation; however the 
estimated cross-section for mini black hole production is generally
too small to affect the majority of EAS. 

\item The energy threshold $E_{\rm th}$ for the new effect lies
between $10^{17.5}-10^{18.3}$ eV ( CM energy 25 - 60 TeV),
within reach of any next-generation accelerators.  

\item The compositon of the extragalactic cosmic ray
population is light and stable with energy. This could, in principle, be
independently tested electromagnetically, for example by propagation
studies in the Galactic magnetic field, provided that an accurate tomographic
mapping for the latter becomes available. Should such a confirmation be
made, it would 
  {\em necessitate } the existence of new physics around 50 TeV. Another central
  factor in such efforts is good statistics at the highest
  energies. Next-generation cosmic-ray experiments will thus play a
  key role in our ability to use UHECR as probes of new physics. 
\end{enumerate}

\acknowledgements{}
We thank P. Sphicas, N. Kylafis, and K. Tassis for useful
discussions, and M. Unger for helpful feedback on an early version of
this work. TT wishes to thank CERN-TH for their hospitality during the
late stages of this work.

\end{document}